# Quantitative modeling of complex molecular response in coherent cavity-enhanced dual-comb spectroscopy

Adam J. Fleisher,* David A. Long, and Joseph T. Hodges

*Material Measurement Laboratory, National Institute of Standards and Technology, 100 Bureau Drive, Gaithersburg, Maryland 20899, U.S.A.*

*Correspondence to:  adam.fleisher@nist.gov

**Abstract:**  We present a complex-valued electric field model for experimentally observed cavity transmission in coherent cavity-enhanced (CE) multiplexed spectroscopy (i.e., dual-comb spectroscopy, DCS). The transmission model for CE-DCS differs from that previously derived for Fourier-transform CE direct frequency comb spectroscopy [Foltynowicz et al., *Appl. Phys. B* **110**, 163-175 (2013)] by the treatment of the local oscillator which, in the case of CE-DCS, does not interact with the enhancement cavity. Validation is performed by measurements of complex-valued near-infrared spectra of CO and $CO_2$ by an electro-optic frequency comb coherently coupled to an enhancement cavity of finesse $F$ = 19600. Following validation, we measure the 30012 ← 00001 $^{12}C^{16}O_2$ vibrational band origin with a combined standard uncertainty of 770 kHz (fractional uncertainty of $4 \times 10^{-9}$).

# 1. Introduction

Optical frequency combs (OFCs) [1-4] enable increasingly precise applications of atomic and molecular spectroscopy [5], including primary thermometry [6, 7], optical radiocarbon dating [8], sub-Doppler and Doppler-free spectroscopy [9-26], ultrafast and multidimensional spectroscopy [27-30], survey spectroscopy of molecular ions and cold molecules [31-33], and time-resolved spectroscopy for fundamental chemical kinetics [34-37]. Applied spectroscopies [38] also make novel use of optical frequency combs for actively monitoring greenhouse gas fluxes [39-41], methane for open-path detection [42] and unambiguous source attribution [43] and reactive atmospheric species [44], elucidating the chemical composition of combustion and open flames [45, 46], and for developing nonlinear spectroscopies for medical imaging and diagnosis [47, 48]. In additional, laboratory trace gas sensing is also aided by several fundamental properties of optical frequency combs, naming their high temporal and spatial coherence and high spectral resolution, which enable efficient coupling to an enhancement cavity [49, 50].

Despite the burgeoning list of applications utilizing OFCs, there still exists certain areas of spectroscopy which have proven challenging for combs. One example is the merger of dual-comb spectroscopy [51] with cavity-enhanced direct frequency comb spectroscopy [49, 50], and thus the pairing of high acquisition rates and a high duty cycle with the highest possible sensitivities. This combination has the potential to enable time-resolved spectroscopy with truly zero dead-time. In practice, cavity-enhanced dual-comb spectroscopy (CE-DCS) requires high mutual coherence between three entities: a probe comb, the enhancement cavity, and a second comb serving as a local oscillator (LO). On timescales routinely required to achieve a sufficient signal-to-noise ratio for trace gas detection (≥1 s), this remains technically challenging. Complexity associated with phase locking these three entities has limited CE-DCS using mode-locked lasers to a single and seminal proof-of-principle demonstration [52]. In principle, adaptive sampling and/or real-time phase correction approaches to DCS [53-58] should be amenable to CE-DCS, thus mitigating inherent technical complexity associated with maintaining several high-bandwidth phase-locked servo loops.

Recently, an alternative approach to CE-DCS using electro-optic (EO) frequency combs with a high degree of frequency agility and robustness was demonstrated [59, 60]. Electro-optic frequency combs originating from a single continuous-wave (CW) laser have inherently high mutual coherence, and therefore were readily used to record the CE-DCS spectrum of $CO_2$ [59], as well as the simultaneous spectrum of $CO_2$, CO, $H_2O$, and HDO in synthetic air via the demonstration of coherent CE-DCS, capable of averaging time-domain CE-DCS waveforms for >7200 s [60]. Importantly, CE-DCS with EO frequency combs can be readily extended to a much broader, potentially power-leveled optical bandwidth via any combination of intensity, acousto-optic (AO) and/or EO modulator comb-generation schemes [61-73].

    Here we present a model for the complex-valued (i.e., amplitude and phase) electric field components of an OFC which are transmitted through a high-finesse optical resonator containing an absorbing gas sample. Using a heterodyne detection scheme, the transmitted electric field of the probe OFC is interrogated by and down-converted to radiofrequencies by a second local oscillator (LO) OFC. Unlike other well-established Fourier transform methods in CE comb spectroscopy [74], a full complex-valued electric field model for cavity transmission in CE-DCS has not yet been reported. To validate the model derived herein, we performed CE-DCS on samples of either dilute CO or pure $CO_2$ with EO frequency combs in the near-infrared. The EO frequency combs were generated using dual-drive Mach-Zehnder modulators (DMZs) chosen for their relatively flat power spectrum [75, 76] and sufficient bandwidth for precision molecular spectroscopy [57, 59]. Following model validation, we implemented quantitative CE-DCS measurements of the 30012 ← 00001 $^{12}C^{16}O_2$ vibrational band (38 rovibrational transitions) near λ = 1575 nm, determining the vibrational band origin frequency with a combined standard uncertainty of 770 kHz (fractional uncertainty of 4 × 10$^{-9}$).

## 2. Model for complex cavity transmission

### 2.1. General cavity transmission expressions

Many of the concepts introduced in this section are illustrated in Fig. 1. For a two-mirror optical resonator of length $L$, the generalized complex round-trip gain factor $\tilde{g}_{rt}$ as a function of angular optical frequency $\omega = 2\pi\nu$ is defined by Siegman [77] in terms of the mirror electric field amplitude reflectivity $r_N$, the intracavity single-pass absorption intensity coefficient $\alpha$, and the round-trip phase-shift coefficient $\varphi$ as:

$$\tilde{g}_{\rm rt}(\omega) \equiv r_1(\omega) r_2(\omega) \exp(-\alpha(\omega)L - i\varphi(\omega)). \tag{1}$$

After invoking the conservation of energy relation $1 = r_1(\omega)r_2(\omega) + t_1(\omega)t_2(\omega) + \ell_1(\omega)\ell_2(\omega)$, where $\ell_N(\omega)$ is the electric field losses for mirror $N$, and $t_N(\omega)$ is the electric field transmission for mirror $N$, the complex electric field transmitted through a two-mirror optical resonator is:

$$\tilde{E}_t(\omega) = \tilde{E}_i(\omega) \frac{t_1(\omega)t_2(\omega)}{\sqrt{r_1(\omega)r_2(\omega)}} \frac{\sqrt{\tilde{g}_{\rm rt}(\omega)}}{1-\tilde{g}_{\rm rt}(\omega)}, \tag{2}$$

where $\tilde{E}_i$ is the complex-valued, incident optical electric field.

Substituting Eq. (1) into Eq. (2) and defining the effective intensity coefficients $R = \sqrt{r_1 r_2}$ and $T = \sqrt{t_1 t_2}$ for the mirror pair gives,

$$\tilde{E}_t(\omega) = \tilde{E}_i(\omega) \frac{T(\omega) \exp[-\alpha(\omega)L/2 - i\varphi(\omega)/2]}{1 - R(\omega)\exp[-\alpha(\omega)L - i\varphi(\omega)]}. \tag{3}$$

Equation (3) is the generalized complex expression for the transmitted electric field as defined by the properties of the optical resonator and the absorbing intracavity medium.

In CE-DCS, the frequency of the transmitted optical electric field is down-converted into the radiofrequency (RF) domain by combination with an LO electric field $\tilde{E}_{LO}$ which produces a suite of heterodyne beat signals which are measured with a fast photodetector. The resulting interferogram contains both the amplitude and phase of the multiheterodyne signal $\tilde{E}_t \tilde{E}_{LO}^*$. The normalized intensity ($I_t/I_0$) and phase ($\Phi$) spectrum of the cavity transmission is defined in Eqs. (4) and (5).

$$\frac{I_t}{I_0} = \frac{|\tilde{E}_t \tilde{E}_{LO}|}{|\tilde{E}_0 \tilde{E}_{LO}^*|} \tag{4}$$

$$\Phi = \tan^{-1}\left[\frac{\Im(\tilde{E}_t \tilde{E}_{LO}^*)}{\Re(\tilde{E}_t \tilde{E}_{LO}^*)}\right] - \tan^{-1}\left[\frac{\Im(\tilde{E}_0 \tilde{E}_{LO}^*)}{\Re(\tilde{E}_0 \tilde{E}_{LO}^*)}\right] \tag{5}$$

Above, $\tilde{E}_0$ is the transmitted electric field absent intracavity absorbers (i.e., intracavity vacuum) and other residual phase shifts.

$$\tilde{E}_0(\omega) = \tilde{E}_i(\omega)\frac{T(\omega)}{1-R(\omega)} \qquad (6)$$

Finally, the general expression for the normalized, complex-valued electric field transmission is:

$$\frac{\tilde{E}_t}{\tilde{E}_0} = \frac{1-R(\omega)}{\sqrt{R(\omega)}}\frac{\sqrt{\tilde{g}_{\text{rt}}(\omega)}}{1-\tilde{g}_{\text{rt}}(\omega)} = \frac{[1-R(\omega)]\exp[-\alpha(\omega)L/2-i\varphi(\omega)/2]}{1-R(\omega)\exp[-\alpha(\omega)L-i\varphi(\omega)]}. \qquad (7)$$

In practice, spectral normalization can be performed in one of several ways. The spectrum of $\tilde{E}_0(\omega)$ can be recorded with the enhancement cavity at vacuum pressures, or by diverting the incident comb $\tilde{E}_i(\omega)$ around the optical cavity and through a variable attenuator (as was done here). The exact choice of a normalization procedure will influence the definition of $\tilde{E}_0(\omega)$. As defined in Eq. (6), $\tilde{E}_0(\omega)$ is approximated by a scaled version of $\tilde{E}_i(\omega)$. Importantly, the definition of $\tilde{E}_0(\omega)$ in Eq. (6) assumes no phase shifts related to cavity-comb coupling (including cavity mirror dispersion). Put another way, Eq. (6) assumes that all comb teeth are perfectly matched to the center of their unique optical cavity resonances in the absence of molecular absorption. As we choose to bypass the optical cavity and scale $\tilde{E}_i(\omega)$, our complex transmission spectra will include contributions from cavity-comb dispersion (Section 2.3.2).

*2.2. Limiting cases – weak/strong resonant intracavity absorption*

For the high-reflectivity case considered here, Eq. (7) exhibits a series of widely spaced narrow resonances that can be treated individually. Neglecting the single-pass term in the numerator, assuming that both $\alpha L$ and $\varphi$ are much less than unity and linearizing the exponential term in the denominator, then the transmitted signal in the vicinity of an individual cavity resonance reduces to the complex-valued Lorentzian function

$$\frac{\tilde{E}_t(\omega)}{\tilde{E}_0} = \frac{1}{\left(1+\frac{F}{\pi}\alpha(\omega)L\right)+i\frac{F}{\pi}\varphi(\omega)}, \qquad (8)$$

where $F = \frac{\pi}{1-R}$ is the empty-cavity finesse. Equation (8) has a normalized magnitude that is closely approximated by

$$\frac{I_t(\omega)}{I_0} = 1 - \frac{F}{\pi}\alpha(\omega)L, \qquad (9)$$

at the on-resonance condition corresponding to $\varphi = 0$. This result is the familiar expression for the cavity-enhanced transmission spectrum of a heterodyne signal in the low-absorption limit. More generally, for off-resonance excitation of a given cavity mode, the normalized magnitude, $\frac{I_t(\omega)}{I_0}$ and phase, $\Phi(\omega)$, of the complex-valued heterodyne transmission signal are given by

$$\frac{I_t(\omega)}{I_0} = \frac{1}{\sqrt{\left(1+\frac{F}{\pi}\alpha(\omega)L\right)^2+\left(\frac{F}{\pi}\varphi(\omega)\right)^2}}, \qquad (10)$$

and

$$\Phi(\omega) = -\tan^{-1}\left(\frac{\frac{F}{\pi}\varphi(\omega)}{1+\frac{F}{\pi}\alpha(\omega)L}\right). \qquad (11)$$

Note that the on-resonance cavity enhancement factor $F/\pi$ for CE-DCS appearing is half that found in Fourier transform (FT) frequency comb spectroscopy [74]. Also, inspection of Eqs. (9) and (11), illustrates that this on-resonance factor amplifies both the absorbance and phase-shift coefficients equally.

In the limit of high absorption where $F\alpha(\omega)L \gg 1$, changes in the round-trip phase shift are dominated by absorption-induced dispersion in the cavity medium (see Sec. 2.3.1. below). For cases where absorption corresponds to discrete quantum transitions or a linear combination thereof, there is a simple relationship between the phase of the heterodyne signal and that of the complex-valued line shape $\tilde{g}(\omega)$. Assuming perfect matching between the interrogating comb and empty-cavity modes, the phase of the transmission signal becomes

$$\Phi(\omega) = -\tan^{-1}\left(\frac{\varphi(\omega)}{\alpha(\omega)L}\right) = \tan^{-1}\left(\frac{\Im[\tilde{g}(\omega)]}{\Re[\tilde{g}(\omega)]}\right), \tag{12}$$

where $\Re[\tilde{g}(\omega)]$ and $\Im[\tilde{g}(\omega)]$ represent the real (absorptive part) and imaginary component (dispersive part), respectively, of $\tilde{g}(\omega)$. In this limit, the phase of the transmitted heterodyne signal equals the phase of the complex-valued line shape. Moreover, for the more general case corresponding to an arbitrary amount of absorption and to non-zero mismatch between the comb and empty-cavity mode spacing, the complete solution (discussed below) remains sensitive to the complex-valued line shape. The ability to interrogate both the real and imaginary components of the line shape contrasts with conventional cavity-enhanced spectroscopy methods which are only sensitive to absorption. Applications of such complex-valued line profile measurements include the study of advanced line shapes [78], as well as broadband effects like non-Lorentzian behavior in the far-wings, line mixing and collisional induced absorption [79, 80].

### 2.3. The round-trip phase shift coefficient

The round-trip phase shift coefficient $\varphi(\omega)$ is treated here as the sum of two parts: resonant molecular dispersion [$\varphi_\mathrm{m}(\omega)$] and cavity-comb dispersion [$\varphi_\mathrm{c}(\omega)$], each defined relative to the grid of optical frequencies established by the laser comb. The derivation of $\varphi(\omega)$ presented here makes use of the prior works of Thorpe et al. [81] and Foltynowicz et al. [74] as well as the following textbooks and book chapters: Siegman [77], Yariv [82], Lehmann [83], Boyd [84], and Nagourney [85].

### 2.3.1. Resonant molecular dispersion

The round-trip phase-shift coefficient for resonant molecular dispersion is defined as $\varphi_\mathrm{m}(\omega) = \varphi'_\mathrm{m}(\omega) - \varphi^0_\mathrm{m}(\omega)$, where $\varphi'_\mathrm{m}$ and $\varphi^0_\mathrm{m}$ are the perturbed and unperturbed optical phases at each cavity mode. Absent mirror dispersion (Section 2.3.2), the unperturbed optical phases are defined as $\varphi^0_\mathrm{m}(\omega) = \omega t_\mathrm{rt}$, where $t_\mathrm{rt} = 1/\nu_\mathrm{FSR}$ is the unperturbed cavity round-trip time (equal to the inverse of the cavity free spectral range $\nu_\mathrm{FSR}$). Defining $\nu_{FSR} = \nu^0_q/q$, where $\nu^0_q$ is the nominal optical frequency of longitudinal mode order $q$, and returning to angular optical frequency, we write $\varphi_\text{m}^0(\omega) = 2\pi q\omega/\omega_q^0$.

The angular frequencies of the perturbed cavity modes are $\omega_q = \omega_q^0\sqrt{1 - \chi'(\omega)}$, where $\chi'(\omega)$ is the real part of the complex linear susceptibility $\tilde{\chi}(\omega)$. If $|\chi'(\omega)| \ll 1$, the perturbed cavity mode angular frequencies are well-approximated by $\omega_q = \omega_q^0[1 - \chi'(\omega)/2]$. This expression for $\omega_q$ yields $\varphi_\text{m}'(\omega) = 2\pi q\omega/\{\omega_q^0[1 - \chi'(\omega)/2]\}$ for the perturbed optical phases. Using the above expressions for $\varphi_\text{m}^0(\omega)$ and $\varphi_\text{m}'(\omega)$, we arrive at an expression for the round-trip phase-shift coefficient for resonant molecular dispersion in Eq. (10).

$$\varphi_\text{m}(\omega) = 2\pi \frac{\omega\chi_s'(\omega)/2}{\omega_\text{FSR}} = 2\pi \frac{\Delta\omega_m}{\omega_\text{FSR}} \tag{13}$$

In Eq. (13), $\Delta\omega_m = \frac{\omega\chi_s'(\omega)}{2}$ and the cavity free spectral range in angular units is $\omega_\text{FSR} = 2\pi\nu_\text{FSR}$. Note that $\tilde{\chi}(\omega)$ could be a linear combination of linear susceptibilities from any number of discrete molecular (or atomic) transitions.

The real part of the molecular susceptibility $\chi'(\omega) = c^2 n_\text{a} S_\text{int}\Im[\tilde{g}(\omega)]/\omega = c\phi(\omega)/\omega$, where $\phi(\omega) = cn_\text{a}S_\text{int}\Im[\tilde{g}(\omega)]$ is the familiar molecular dispersion coefficient in dimensions of inverse length, $n_\text{a} = p_\text{a}/(k_B T)$ is the number density of absorbers, $p_\text{a}$ is the partial pressure of the absorbers, $k_B$ is the Boltzmann constant, $T$ is the sample temperature, and $S_\text{int}$ is the transition line intensity in dimensions of area/(wave number × molecule). Therefore, Eq. (10) can be written in terms of $\phi(\omega)$ as

$$\varphi_\text{m}(\omega) = \phi(\omega)L. \tag{14}$$

*2.3.2. Cavity-comb dispersion*

Dispersion in an optical cavity is conveniently expressed as an expansion of the propagation constant $k(\omega)$ about some angular frequency $\omega_0$ as

$$k(\omega) = k_0 + \left(\frac{\partial k}{\partial \omega}\right)_{\omega_0}(\omega - \omega_0) + \frac{1}{2}\left(\frac{\partial^2 k}{\partial \omega^2}\right)_{\omega_0}(\omega - \omega_0)^2 + \cdots. \tag{15}$$

In this section, we define propagation constant expansions for both the cavity [$k_{\text{cav}}(\omega)$] and the comb [$k_{\text{comb}}(\omega)$], ultimately revealing the cavity-comb coupling dispersion term $\varphi_c(\omega) = 2L[k_{\text{cav}}(\omega) - k_{\text{comb}}(\omega)]$. The zeroth-order expansion coefficients, $k_0$, for the cavity and the comb are $k_{0,\text{cav}} = (\omega_0 - \delta\omega_{\text{PDH}})/c$ and $k_{0,\text{comb}} = \omega_0/c$, respectively, where $\omega_0$ is the angular optical frequency of the cavity mode to which the CW comb seed laser is locked, and $\delta\omega_{\text{PDH}}$ is any offset in the angular frequency of the Pound-Drever-Hall (PDH, [86]) locked comb tooth from the center of the reference cavity mode.

The first-order expansion coefficient $\partial k/\partial \omega$ in Eq. (12) is the inverse group velocity ($1/v_g$). When local changes in $n(\omega)$ are small (i.e., $\omega \partial n(\omega)/\partial \omega \ll 1$, assuming $n(\omega) = 1$), $v_g \approx c/n$. This assumption is consistent with the general assumption that the propagation constant depends weakly on the angular frequency, and therefore allows for the expansion in Eq. (12). In this limiting case we can express the group velocity of the cavity ($v_{g,\text{cav}}$) and the group velocity of the comb ($v_{g,\text{comb}}$) in terms of the cavity free spectral range ($v_{g,\text{cav}} = 2L\nu_{\text{FSR}}$) and the best-match comb repetition rate ($v_{g,\text{comb}} = 2L\nu_{\text{rep}}$).

The second-order expansion coefficient $\partial^2 k/\partial \omega^2$ is the group velocity dispersion (GVD), equal to $\partial(1/v_g)/\partial \omega$. Note that the GVD is equal to the group delay dispersion (GDD) $D_\omega$ divided by the round-trip path length $2L$. Gathered together, these terms yield two unique expressions for the cavity and the comb wave numbers, respectively.

$$k_{\text{cav}}(\omega) \approx \frac{\omega_0 - \delta\omega_{\text{PDH}}}{c} + \frac{1}{2L\nu_{\text{FSR}}}(\omega - \omega_0) + \frac{D_\omega}{4L}(\omega - \omega_0)^2 \qquad (16)$$

$$k_{\text{comb}}(\omega) \approx \frac{\omega_0}{c} + \frac{1}{2L\nu_{\text{rep}}}(\omega - \omega_0) \qquad (17)$$

The round-trip phase shift coefficient for cavity-comb dispersion is $\varphi_c(\omega) = 2L[k_{\text{cav}}(\omega) - k_{\text{comb}}(\omega)]$. Defining the frequency axis reference frame so that $\omega_0$ is the angular frequency of the PDH locked comb tooth, we write $\omega - \omega_0 = \beta\omega_{\text{rep}}$, where $\omega_{rep} = 2\pi\nu_{rep}$ and $\beta$ is a signed integer identifying each unique comb tooth relative to the PDH lock point (see also Section 2.4). Using the above expressions, and neglecting the GDD term in $k_{cav}(\omega)$ over the EO comb bandwidth (an assumption justified in the following paragraph) we arrive at the following equation:

$$\varphi_c(\omega) = 2\pi\left(\frac{\beta\Delta_c - \delta\omega_{\text{PDH}}}{\omega_{\text{FSR}}}\right) = 2\pi\frac{\Delta\omega}{\omega_{\text{FSR}}}, \qquad (18)$$

where $\Delta_c = \omega_{\text{FSR}} - \omega_{\text{rep}}$, and $\Delta\omega = \beta\Delta_c - \delta\omega_{\text{PDH}}$ is the apparent shift in angular frequency at each cavity mode relative to its unique comb tooth. Note that Eq. (18) and Eq. (13) are identical, and reveal an intuitive result: the phase-shift coefficient is simply described as being proportional to the frequency shift at each cavity mode divided by the cavity free spectral range.

For coupling femtosecond laser pulses to enhancement cavities, cavity mirrors with near-zero GDD are desirable. As an example, we see from Fig. 3 of [81] that zero-GDD mirrors can be engineered to have excursions in $|D_\nu| = |D_\omega|/2\pi < 10$ fs$^2$ over an optical bandwidth of 40 nm. For the purposes of estimating the influence of mirror GDD on $\Delta\omega$, we assume $D_\nu = 10$ fs$^2$ (normal dispersion) and calculate $\varphi_c^{(2)}(\omega)$ (where the superscript (2) indicates the use of only the second-order expansion coefficient) at the largest optical detuning from the PDH lock point used herein: $15\nu_{\text{rep}} \approx 3$ GHz. The resulting phase shift is $\varphi_c^{(2)}(\omega) = 1.15 \times 10^{-8}$ rad. Rearranging Eq. (15), we estimate $\Delta\nu = \Delta\omega/(2\pi) = 0.37$ Hz, a frequency shift due to GDD which is much less than the cavity line width of $\delta_{\text{cav}} \approx 10$ kHz. At a mode order of $m = 2500$ (optical detuning of 0.5 THz with $\nu_{\text{rep}} = 203$ MHz), the frequency shift due to a GDD of 10 fs$^2$ is virtually equivalent to $\delta_{\text{cav}}$. Therefore, quantitative modeling of CE-DCS performed with broadband frequency combs will require the inclusion of GDD. (The prerequisite inclusion of GDD has already been alluded to [87] and ultimately exquisitely controlled [88] in the early works on CE spectroscopy using mode-locked lasers and non-DCS methods.) Broadband EO combs [60, 67, 72] provide a unique opportunity to interrogate GDD in a mode-resolved fashion because the CW laser which seeds the probe EO comb generator can be tightly phased-locked to a unique mode of the optical cavity (as was done here and in the first demonstrations of CE-DCS [59] and coherent CE-DCS [60] using EO combs).

Finally, we note that for the general case of locking a broadband comb to a cavity with near-zero GDD at the lock-point wavelength, the next higher-order expansion term, proportional to $\partial D_\omega/\partial\omega$ [which scales as $(\omega - \omega_0)^3$], may not be negligible compared to the GDD term (second-order expansion coefficient).

*2.4. Model expression for spectral fitting*

The CE-DCS demonstrated here using EO frequency combs achieves cavity-comb coupling via a PDH phase lock of a single comb mode to a single resonator mode. For spectral analysis, it is therefore useful to shift the reference frame of Eqs. (1-6) from absolute angular frequency to a relative frequency axis with the locking point as the reference frequency (as was introduced in Section 2.3.2.). The individual comb modes transmitted by the resonator are then identified by their unique signed mode number $\beta$, where $\beta = 0$ at the PDH locking point.

By substitution of $\varphi = \varphi_\mathrm{m} + \varphi_\mathrm{c}$ derived in Section 2.3 into Eq. (1), we arrive at the following expression for the normalized transmitted electric field.

$$\frac{\tilde{E}_t}{\tilde{E}_0} = \frac{[1-R(\beta)]\exp\left\{-\frac{\alpha(\beta)L}{2}-\frac{i[\phi(\beta)-\phi(0)]L}{2}-\frac{i\pi(\beta\Delta_{c,\nu}-\delta_{\mathrm{PDH},\nu})}{\nu_{\mathrm{FSR}}}\right\}}{1-R(\beta)\exp\left\{-\alpha(\beta)L-i[\phi(\beta)-\phi(0)]L-\frac{i2\pi(\beta\Delta_{c,\nu}-\delta_{\mathrm{PDH},\nu})}{\nu_{\mathrm{FSR}}}\right\}} \quad (19)$$

The full model is obtained by substitution of Eq. (19) into Eqs. (4-5) with the addition of a complex-valued linear baseline. For the data analysis reported below, we fit this model to the measured multi-heterodyne transmission spectrum.

## 4. Experimental validation of the transmission model

The complex transmission model presented in Section 2 contains the phase shift parameter $\varphi$ which describes the mismatch between the comb mode spacing and the cavity free spectral range ($\Delta_{c,\nu} = \Delta_c/2\pi = \nu_{\mathrm{rep}} - \nu_{\mathrm{FSR}}$) and the PDH locking offset parameter $\delta_{\mathrm{PDH},\nu} = \delta_{\mathrm{PDH}}/2\pi$. The value of $\Delta_{c,\nu}$ is non-zero if the cavity length drifts 1) from the time that $\nu_{\mathrm{FSR}}$ was last measured, 2) between successive spectral acquisitions, or 3) if $\nu_{\mathrm{rep}}$ of the probe comb is manually detuned from $\nu_{\mathrm{FSR}}$. The value of $\delta_{\mathrm{PDH},\nu}$ is non-zero if the PDH locking electronics introduce a DC offset (potentially when the molecular dispersion at the PDH reference mode is significant).

*4.1. Influence of $\Delta_{c,\nu}$ on the spectrum of carbon monoxide (CO)*

Figures 4 and 5 show the transmission and phase spectrum of 10 kPa (75 Torr) of NIST Standard Reference Material (SRM®) 2637a, CO in $N_2$ (CO concentration of 2472.8 µmol/mol ± 4.2 µmol/mol), recorded with intentional detunings of $\Delta_{c,\nu} = -0.5$ kHz (Fig. 4, approximately $-\delta_{cav}/20$) and $\Delta_{c,\nu} = -2$ kHz (Fig. 5, approximately $-\delta_{cav}/5$), respectively, where $\delta_{cav} = \nu_{FSR}/F$ is the cavity line width. The line center and line strength of the CO transition are $\nu_0 = 6364.767\,625$ cm$^{-1}$ and $S_{int} = 1.551 \times 10^{-23}$ cm/molecule [89]. In Figs. 4 and 5, normalized transmission and phase spectra are plotted in the top and bottom panels, respectively. For a given complex cavity transmission spectrum, we fitted both the transmission and phase spectra simultaneously using Eq. (19) as well as Eqs. (4-5) to retrieve the fitted detuning $\Delta_{c,\nu}$. The fitted models (black lines) reproduced the experimental data points (black circles) well, as evidenced by the absence of systematic structure in the fitted residuals. The retrieved values of $\Delta_{c,\nu} = -456$ Hz and $\Delta_{c,\nu} = -1.92$ kHz are in good agreement with the intended detunings of $\Delta_{c,\nu} = -500$ Hz and $\Delta_{c,\nu} = -2$ kHz. We also observed a fitted value of $\delta_{PDH,\nu} \neq 0$, most likely from an unintended consequence of optimizing the cavity throughput via adjustment of the PDH locking conditions (see Section 4.2).

Also plotted in Figs. 4 and 5 are additional simulations of the transmission and phase spectra. For the blue dashed lines, we reverse the sign of the fitted $\Delta_{c,\nu}$ to be $-\Delta_{c,\nu}$; for the red dotted lines we fix $\Delta_{c,\nu} = \delta_{PDH,\nu} = 0$. The red dotted lines, with $\Delta_{c,\nu} = \delta_{PDH,\nu} = 0$, correspond to the hypothetical transmission and phase spectrum absent round-trip phase shifts associated with the cavity-comb coupling. The additional simulations illustrate the sensitivity of our multiplexed fit to the round-trip phase shift parameter $\varphi$, which we uniquely interrogate in this experiment with full comb tooth (cavity mode) resolution. Importantly, we are also measuring these round-trip phase parameters relative to a single cavity mode with fixed phase, i.e. the PDH locking point with no optical detuning.

*4.2. Influence of $\delta_{PDH,\nu}$ on the spectrum of carbon dioxide ($CO_2$)*

Examples of pure $CO_2$ spectra with near-zero values of $\Delta_{c,\nu}$ and large non-zero values of $\delta_{PDH,\nu}$ are shown in Fig. 6. For this series of complex spectra, cavity throughput (i.e., transmitted optical power) was optimized while the seed laser for the probe comb was locked to an enhancement cavity mode experiencing significant molecular dispersion. By maximizing probe comb throughput, the optical frequency of the PDH-locked seed laser was detuned from the center of the cavity mode with $\beta = 0$, resulting in $\delta_{\text{PDH},\nu} \neq 0$. This condition was achieved by adjusting the zero-crossing of the PDH error signal until a maximum in the down-converted probe comb interferogram was observed. Values of $|\delta_{\text{PDH},\nu}|$ nearly equal to half the empty-cavity mode line width are achieved, although we note that the local mode line width is significantly broadened by the presence of molecular absorption at the lock point. For the spectra in Fig. 6 (60 s acquisition time), the largest fitted value of $|\Delta_{c,\nu}|$ was observed for a $CO_2$ pressure of 39 Pa (black circles), where $\Delta_{c,\nu} = -91(9)$ Hz, followed by $\Delta_{c,\nu} = 80(6)$ Hz for the lowest pressure of 5.6 Pa $CO_2$ (red diamonds) and $\Delta_{c,\nu} = 33(25)$ Hz for the intermediate pressure of 12 Pa $CO_2$ (blue squares).

## 5. Applications

*5.1. Absolute transition frequencies of carbon dioxide ($CO_2$)*

To apply the validated CE-DCS model, we measured absolute frequencies for 38 distinct rovibrational transitions within the 30012 ← 00001 $^{12}C^{16}O_2$ band centered at $\lambda = 1575$ nm. Transition frequencies for this band have previously been measured by FARS cavity ring-down spectroscopy (CRDS) with absolute uncertainties between 30 kHz and 800 kHz [30]. The resulting FARS-CRDS fitted accuracy for the 30012 ← 00001 band origin ($G_\nu$) was 59 kHz, or equivalently a fractional uncertainty of $\sigma_{G_\nu}/G_\nu = 3 \times 10^{-10}$ [90].

Representative transmission (top left panel) and phase (bottom left panel) CE-DCS spectra of $CO_2$ are plotted in Fig. 7. The CE-DCS of a high-purity, isotopically enriched $CO_2$ gas sample (pressure $p = 13$ Pa, temperature $T = 296$ K, $^{12}C$ isotopic purity 0.997 6), spanning nearly 2 THz, was measured in a line-by-line, multiplexed fashion. For each rovibrational transition, 60 multiplexed spectra (30 probe comb lines per spectrum, spaced by $\nu_{\text{rep}} = \nu_{\text{FSR}} \approx 203.085$ MHz) were recorded in 2 s of integrated time per spectrum, and at an average duty cycle of $t_{\text{int}}/t_{\text{total}} = 60\%$. Given lessons learned in Section 4, a slow integrator feeding back to the ECDL cavity lenght was added to the PDH servo to minimize DC offsets in the PDH phase locking servo loop.

Individual interferograms were coherently averaged for 1 s using fast acquisition software and then analyzed in the time domain to retrieve the amplitude and phase at each comb mode [60]. The transmission and phase spectra were modeled using Eq. (19) and Eqs. (4-5), and fitted using a non-linear least-squared algorithm. Transition frequencies, fitted relative to the PDH locked carrier frequency, were ultimately referenced to a commercial self-referenced OFC as illustrated in Fig. 2. The $f_{\text{ceo}}$ of the reference comb was stabilized via an $f$-$2f$ interferometer, and the reference comb $f_{\text{rep}}$ was stabilized to the same Cs clock signal used to stabilize both EO frequency comb generation and coherent DCS interferogram collection.

Following the consecutive acquisition of 60 multiplexed spectra, the ECDL was rapidly tuned to the next rovibrational transition and multiplexed spectral acquisition was resumed. Because molecular dispersion induces asymmetries in the observed transmission spectra, we recorded the entire 30012 ← 00001 band with two different constant frequency offsets between the PDH locked carrier comb mode and the expected molecular transition frequency of ±1.2 GHz, or approximately ±4 cavity modes. Despite the intentional frequency offset, which was intended to minimize the influence of molecular dispersion on the fitted parameter $\Delta_{c,v}$, phase offsets are still clearly observable in the bottom panels of Fig. 7. In the bottom right panel of Fig. 7, the black and the blue fitted R14e transitions exhibit nearly equal in magnitude but opposite in sign phase shifts relative to the empty-cavity mode phases (i.e., $\phi = 0$). This is expected, and fully captured at every probe comb tooth by our complex cavity transmission model without a priori knowledge of the molecular or mirror dispersion.

### 5.2. Accurate spectroscopic parameters for the 30012 ← 00001 $^{12}C^{16}O_2$ band

Measured transition frequencies (Section 5.1) are plotted in Fig. 8 vs. *m*, where here *m* = –*J"* for the P-branch (Δ*J* = *J'* – *J"* = –1) and *m* = *J"* + 1 for the R-branch (Δ*J* = +1), where *J"* and *J'* are the lower- and upper-rotational state quantum numbers. The absolute transition frequencies plotted in Fig. 8 are the weighted average of 60 individual multiplexed spectra. Fitted residuals from the spectroscopic model (black line, top panel) are plotted in the bottom panel of Fig. 8. The fitted spectroscopic model yielded values for the 30012 ← 00001 vibrational band origin frequency $G_v$ along with the upper-state inertial parameters $B_v'$, $D_v'$, and $H_v'$ [91]. For the fitted model, the lower-state inertial parameters were fixed at the values reported in Ref. [90].

Experimental error bars ($\pm 1\sigma$ uncertainty) plotted in the bottom panel of Fig. 8 are the average standard deviation from the two 60-spectra ensembles, with two exceptions (P10e and R12e, with *m* = −10 and *m* = 13) where absolute transition frequencies were only measured at a single detuning PDH locking point from the molecular transition frequency due to a loss in $f_{\text{ceo}}$ stabilization of the self-referenced OFC during CE-DCS data acquisition.

Fitted values of $G_v$, $B_v'$, $D_v'$, and $H_v'$ are listed in Table 1. By fitting 38 unique absolute transition frequencies across the 30012 ← 00001 band, we report $G_v$ with a combined standard uncertainty of $\sigma_{G_v} = 770$ kHz and a relative uncertainty of $\sigma_{G_v}/G_v = 4 \times 10^{-9}$. All CE-DCS parameters measured here and listed in Table 1 exhibit $\pm 2\sigma$ agreement with those measured by FARS-CRDS [90].

## 6. Conclusions

We derive, validate and apply a complex transmission model for coherent cavity-enhanced dual-comb spectroscopy (CE-DCS), and use the model to measure the 30012 ← 00001 $^{12}C^{16}O_2$ vibrational band origin with a standard uncertainty of 770 kHz. We also report low-uncertainty measurements of the upper-state inertial parameters of $CO_2$ which agree within $\pm 2\sigma$ of the literature values. With a total integration time of 76 s, we recorded virtually an entire vibrational band of $CO_2$ covering an optical frequency range of nearly 2 THz.

Uniquely, coherent CE-DCS enables the multiplexed interrogation of enough enhancement cavity modes to retrieve the full complex (amplitude and phase) molecular response of individual rovibrational transitions without laser scanning, thus "freezing" potential sources of systematic uncertainty. With high acquisition rates (up to $2 \times 1/\Delta\nu_{\text{rep}}$) and a high-throughput data stream amenable to actively controlled coherent averaging [60], post-processing [57], and/or adaptive sampling [58], deep averaging of fitted spectroscopic parameters is possible, and therefore so is high precision. Although intensity measurements of cavity transmission modes can also measure complex molecular response (via mode position and line width) [92-95], those techniques each require laser scanning, and therefore are inherently susceptible to complications arising from temperature drifts, mechanical and acoustic vibrations, polarization and birefringence drifts, and parasitic etalons. The targeted, accurate multiplexed measurements reported herein are readily extended to continuous coverage over bandwidths of at least 2 THz by leveraging the rapid-scanning capabilities of modern continuous-wave (CW) lasers. For example, external cavity diode lasers are commercially available with tuning rates of up to 20 nm/s (2.4 THz/s at $\lambda = 1575$ nm). With the validated complex transmission model presented here, we envision routinely performing coherent CE-DCS measurements of individual rovibrational transitions without laser scanning, and then acquiring additional targeted rovibrational transitions with rapid and broadband laser agility, thus bridging the gap between precision molecular spectroscopy using CW lasers and broadband frequency combs while leveraging the high optical and data throughput advantages of both types of laser sources.

**Acknowledgements**

This work was funded by the National Institute of Standards and Technology (NIST) Greenhouse Gas Measurement and Climate Research Program.

**Table 1**

**Table 1.** 30012 ← 00001 $^{12}C^{16}O_2$ band origin ($G_v$ in MHz) and upper-state inertial parameters ($B'_v$, $D'_v$ and $H'_v$ in MHz) measured by CE-DCS (this work) compared with literature values measured by FARS-CRDS [90]. Experimental uncertainties ($1\sigma$) are in parentheses. Ground-state inertial parameters were fixed to those reported in Ref. [90].

|  | $G_v$ | $B'_v$ | $D'_v$ ($10^{-3}$) | $H'_v$ ($10^{-9}$) |
|---|---|---|---|---|
| Ref. [90] | 190 303 782.258(59) | 11 585.628 64(39) | −2.942 62(69) | 15.81(32) |
| This work | 190 303 781.28(77) | 11 585.636 7(51) | −2.954 2(85) | 20.6(3.6) |

**Figure 1**

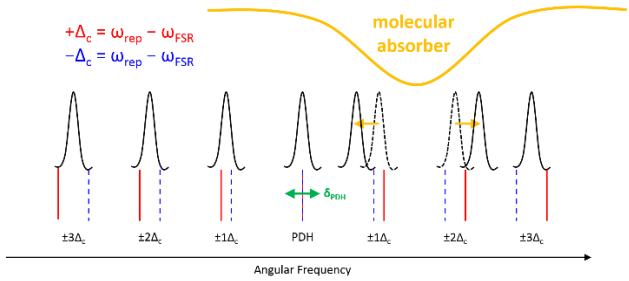

**Fig. 1.** Frequency-domain illustration of comb-cavity coupling in the presence of intracavity absorption. An electro-optic (EO) frequency comb (blue/red lines) is phase locked to a unique cavity mode (black curves) at the points labeled PDH (Pound-Drever-Hall). Each comb tooth has a unique detuning frequency of $\pm\beta\Delta_c$ (see the main text). In the presence of absorption (orange), the EO comb complex transmission spectrum can unambiguously differentiate between $+\Delta_c$ (red) and $-\Delta_c$ (blue dashed) cases. Above, two orange arrows emphasize the local cavity dispersion by dramatically moving the cavity modes from their expected vacuum frequencies (dashed black curves). Finally, the absorption enables the theoretical model to fit any detuning of the PDH locking point $\delta_{\text{PDH}}$ (green double-headed arrow).

**Figure 2**

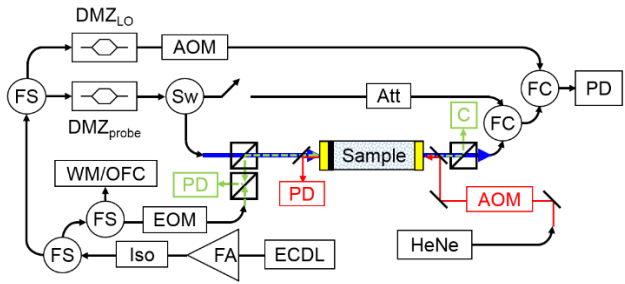

**Fig. 2.** Optical layout for coherent CE-DCS using EO frequency combs. Fiber components are in black, the free-space HeNe laser beam path and optics are in red, the probe comb free-space path is in blue, and the free-space continuous-wave Pound-Drever-Hall locking laser path is in green. Acronyms are defined as follows: AOM, acousto-optic modulator; ECDL, external cavity diode laser; FA, fiber amplifier; Iso, optical isolator; FS, fiber splitter; EOM, electro-optic phase modulator; WM, wavelength meter; OFC, optical frequency comb; PD, photodetector; DMZ, dual-drive Mach-Zehnder modulator; Sw, fiber optic switch; Att, variable optical attenuator; FC, fiber combiner; C, near-infrared camera.

**Figure 3**

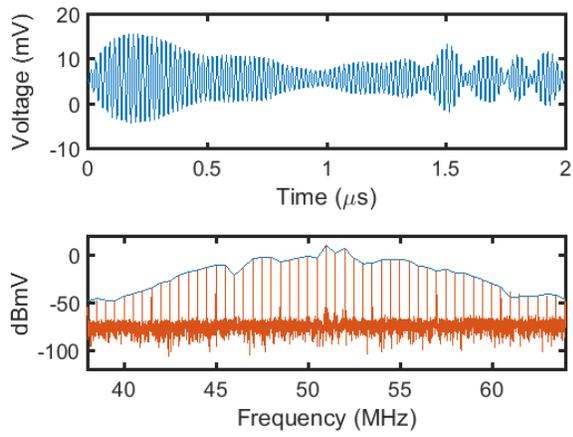

**Fig. 3.** A single time-domain interferogram (period of 2 μs), upper panel, and its corresponding Fourier transform, bottom panel, are shown in blue. In the bottom panel, the Fourier transform of 100 consecutive interferograms is also show in orange.

**Figure 4**

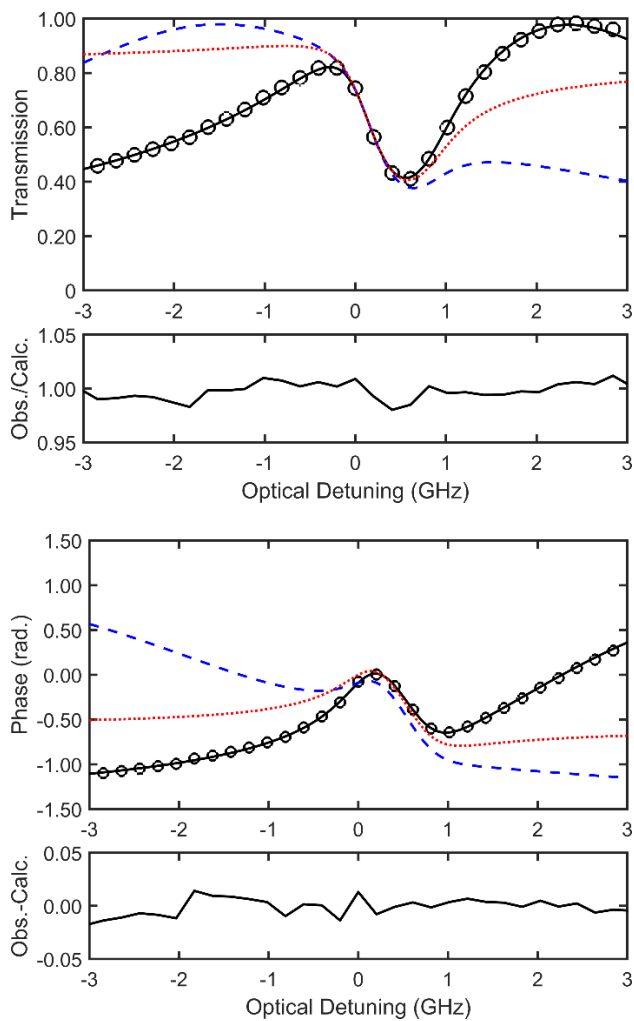

**Fig. 4.** Transmission (top) and phase (bottom) spectrum of 10 kPa of 2472.8 µmol/mol CO in a balance of N$_2$ recorded with an intentional detuning of $f_{\text{rep}}$ from $f_{\text{FSR}}$ of $\Delta_{c,\nu} = -500$ Hz. Experimental data points are plotted in the black circles and the fitted model in the solid black lines. Along with $\Delta_{c,\nu}$, the PDH offset parameter $\delta_{\text{PDH},\nu}$ was also floated in the fitting routine, resulting in fitted values of $\Delta_{c,\nu} = -456(5)$ Hz and $\delta_{\text{PDH},\nu} = -630(50)$ Hz for the black trace. The blue dashed lines show the model calculated at $\Delta_{c,\nu} = +456$ kHz, clearly revealing our ability to unambiguously determine the sign of $\Delta_{c,\nu}$. The red dotted lines are calculated with $\Delta_{c,\nu} = \delta_{\text{PDH},\nu} = 0$, representative of only the cavity-enhanced complex molecular response.

**Figure 5**

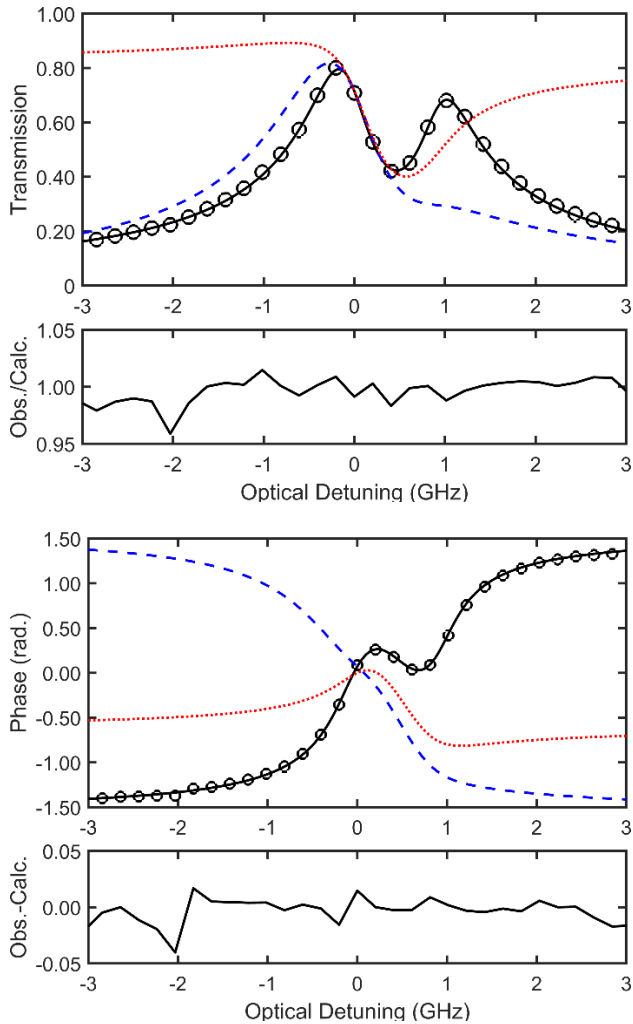

**Fig. 5.** Transmission (top) and phase (bottom) spectrum of 10 kPa of 2472.8 µmol/mol CO in a balance of N$_2$ recorded with an intentional detuning of $f_{\text{rep}}$ from $f_{\text{FSR}}$ of $\Delta_{\text{c},\nu} = -2$ kHz. Experimental data points are plotted in the black circles and the fitted model in the solid black lines. Along with $\Delta_{\text{c},\nu}$, the PDH offset parameter $\delta_{\text{PDH},\nu}$ was also floated in the fitting routine, resulting in fitted values of $\Delta_{\text{c},\nu} = -1.91(6)$ kHz and $\delta_{\text{PDH},\nu} = 510(90)$ Hz for the black trace. The blue dashed lines show the model calculated at $\Delta_{\text{c},\nu} = +1.92$ kHz, again clearly revealing our ability to unambiguously determine the sign of $\Delta_{\text{c},\nu}$. The red dotted lines are calculated with $\Delta_{\text{c},\nu} = \delta_{\text{PDH},\nu} = 0$, representative of only the cavity-enhanced complex molecular response.

**Figure 6**

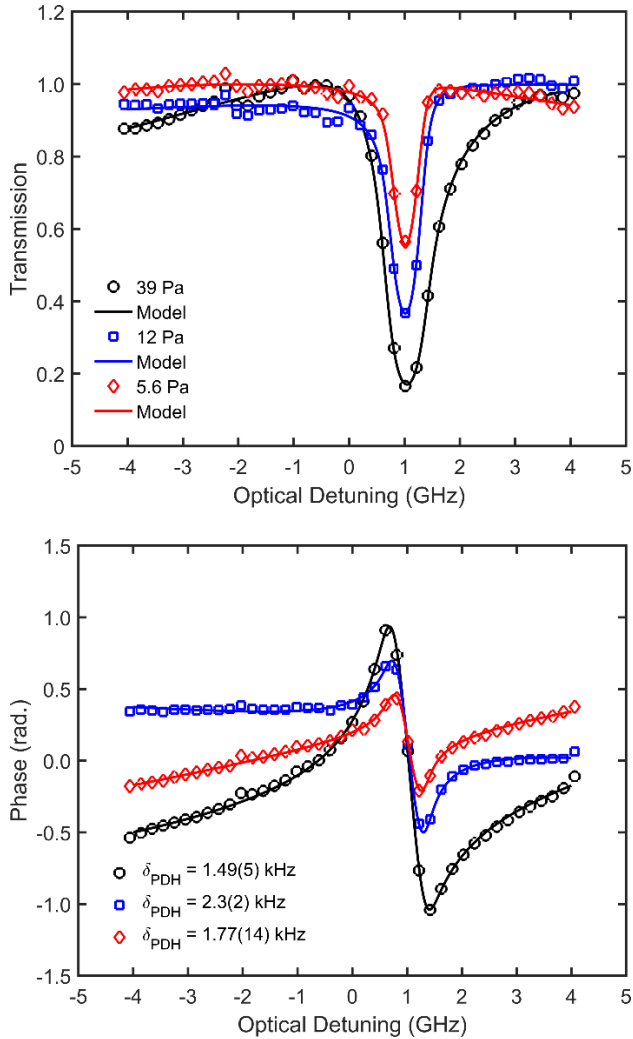

**Fig. 6.** Transmission (top) and phase (bottom) spectrum of pure $CO_2$ at pressures of 5.6 Pa (red), 12 Pa (blue) and 39 Pa (black). The approximate line center and line strength of the transition are $\tilde{v}_0 = 6336.242\,389$ cm$^{-1}$ and $S_{\text{int}} = 1.619 \times 10^{-23}$ cm/molecule [77]. Cavity-comb coupling conditions were optimized for maximum transmission of optical power through the cavity with $\Delta_{c,v} \approx 0$. To achieve these conditions, $\delta_{\text{PDH},v}$ was forced to be large by adjusting the PDH locking conditions. Importantly, the model described in Section 2 (solid lines) is well fitted to the experimental data (open points) in all cases without a priori knowledge of either the magnitude or the sign of $\delta_{\text{PDH},v}$, thus further validating the functionality of the complex transmission model.

**Figure 7**

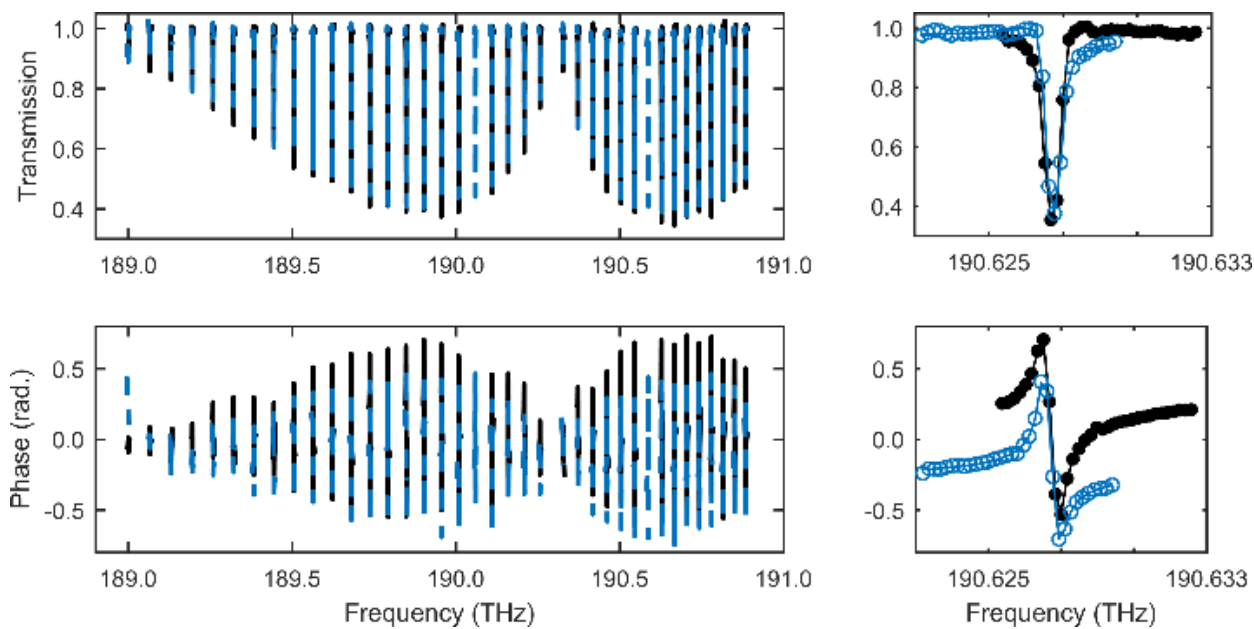

**Fig. 7.** Transmission (top panels) and phase (bottom panels) spectrum of the 30012 ← 00001 $^{12}C^{16}O_2$ band. In the left panels, the band-wide spectrum is shown at two different detunings of the PDH locking point from the individual $CO_2$ transitions frequencies of $\pm 1.2$ GHz (black solid lines and blue dashed lines, respectively). The band-wide spectrum was recorded in a total integration time of 76 s (38 transitions, 2 s per spectrum). In the right panels, spectra of the R14e transition is shown. The experimental data points are plotted as solid black dots (transmission) and open blue circles (phase), while the fitted models are plotted as solid black lines and solid blue lines.

**Figure 8**

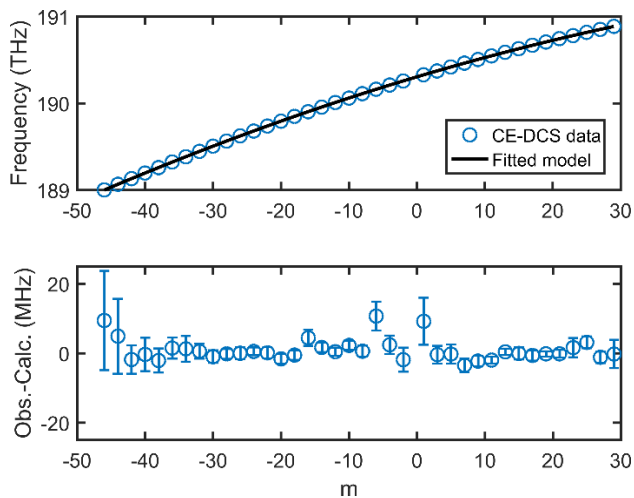

**Fig. 8.** Top panel: Measured absolute transitions frequencies within the 30012 ← 00001 $^{12}C^{16}O_2$ band (blue circles) plotted vs. *m*, along with the fitted spectroscopic model (black line). Bottom panel: Residuals from the fitted model, plotted along with experimental uncertainties (1$\sigma$, standard deviation of the fitted absolute transitions frequencies from 120 individual spectra, each recorded in 2 s of integration time).